\documentclass[aps,prl,twocolumn]{revtex4-1}
\usepackage{amsmath}
\usepackage{amssymb}
\usepackage{natbib}
\usepackage{physics}
\usepackage{hyperref}
\usepackage{graphicx}
\usepackage{color}

\begin{document}
\title{Interaction of Skyrmions and Pearl Vortices in
	Superconductor-Chiral Ferromagnet Heterostructures}
\author{Samme M. Dahir, Anatoly F. Volkov, and Ilya M. Eremin}
\affiliation{Institut f\"ur Theoretische Physik III,
	Ruhr-Universit\"{a}t Bochum, D-44780 Bochum, Germany}
\date{\today }	

\begin{abstract}
	We investigate a hybrid heterostructure with magnetic skyrmions \((\mathrm{Sk})\) inside a chiral ferromagnet interfaced by a thin superconducting film via an insulating barrier. The barrier prevents the electronic transport between the superconductor and the chiral magnet, such that the coupling can only occur through the magnetic fields generated by these materials. We find that Pearl vortices \((\mathrm{PV})\) are generated spontaneously in the superconductor within the skyrmion radius, while anti-Pearl vortices (\(\overline{\mathrm{PV}}\)) compensating the magnetic moment of the Pearl vortices are generated outside of the \(\mathrm{Sk}\) radius, forming an energetically stable topological hybrid structure. Finally, we analyze the interplay of skyrmion and vortex lattices and their mutual feedback on each other. In particular, we argue that the size of the skyrmions will be greatly affected by the presence of the vortices offering another prospect of manipulating the skyrmionic size by the proximity to a superconductor.
\end{abstract}

\maketitle

\date{\today}

Skyrmions (\(\mathrm{Sk}\)) are topologically stable field configurations which have been originally proposed by Tony Skyrme in 1962 in the context of particle physics \cite{skyrme_unified_1962}. The first observation of the so-called chiral skyrmions in chiral ferromagnets in 2009 \cite{muehlbauer_skyrmion_2009}, stimulated an intense research efforts devoted towards the realization of skyrmion-hosting systems suitable for applications. Initially, the appearance of magnetic skyrmions in ferromagnets with Dzyaloshinskii-Moriya interaction \cite{dzyaloshinsky_thermodynamic_1958,moriya_anisotropic_1960} has been predicted in Ref. \cite{bogdanov1989thermodynamically}, refereed to as magnetic "vortices".  Although magnetic skyrmions were identified in single crystals of magnetic compounds with a non-centrosymmetric lattice \cite{muehlbauer_skyrmion_2009,yu_real-space_2010,heinze_spontaneous_2011}, there were more recently observed in ultrathin magnetic films epitaxially grown on heavy metals such as Fe/Ir(111) interfaces \cite{fert_magnetic_2017}. Those interfaces are subject to large Dzyaloshinskii-Moriya interaction (DMI) induced by the broken inversion symmetry at the interface and due to the strong spin-orbit coupling of the neighboring heavy metal atoms.\cite{Heinze_2011}. 

The theory of magnetic skyrmions has been developed in many publications (see for example \cite{bogdanov1989thermodynamically,bogdanov_thermodynamically_1994,bogdanov1995new,rosler_spontaneous_2006,leonov2016,fert_magnetic_2017}). Skyrmions are accompanied by a degree of freedom known as their helicity \(\psi\), which is determined by their spin swirling direction. Their finite extension allows them to move or interact as particles and to be excited at specific dynamical modes making them attractive for spintronics applications. 

At the same time, one of the interesting aspects of the potential development in the field of spintronics is a heterostructures consisting of magnetic and superconducting layers. As a matter of fact, the interplay between superconductivity and ferromagnetism in hybrid structures has received much attention in recent years \cite{Ber_2005,Buz_2005,eschrig2011spin,eschrig2015spin,balatsky_odd_2017}, due to its interest from a fundamental physics viewpoint and also because of improved and new functionality brought about by using superconductors in spintronics \cite{Lin_2015}. Nevertheless, the interaction between topologically non-trivial magnetic inhomogeneity - skyrmions and superconducting vortices remains largely unexplored. Earlier experimental \cite{Sos_2006,Hal_2009,Kal_2011} and theoretical \cite{Volkov_2003,Volkov_2006,Wu2012,Bjornson_2014,Yokoyama_2015,Pershoguba_2016,bobkova_spin_2018} studies concern mostly the Josephson effect in single Josephson SFS junctions with a ferromagnet F containing non-uniform (chiral) magnetic texture, topological classifications and the impurity bound states, induced by the skyrmions. Furthermore, more recent theoretical work discussed the presence of skyrmions in \(\mathrm{F}\)/\(\mathrm{SC}\) heterostructures where a strong proximity effects plays a crucial role \cite{Hals_2016}. The authors used a ballistic limit so that it is yet not clear whether the proposed effects can be observed in a realistic heterostructure.

In this letter we consider a heterostructure consisting of a chiral ferromagnet hosting magnetic skyrmions, which is interfaced by a thin superconducting film \(\mathrm{SC}\) via a thin insulating barrier such that the interaction occurs only via the magnetic fields generated by skyrmions and superconducting vortices. We will show that in this geometry the vicinity of \(\mathrm{Sks}\) induces the so-called Pearl vortices \((\mathrm{PV})\)\cite{pearl_current_1964,gennes_superconductivity_1999} in the \(\mathrm{S}\) film, which affects in turn the \(\mathrm{Sk}\) size and lowers its energy. 
We start by considering the simplest case of a Bloch \(\mathrm{Sk}\) in the presence of a single \(\mathrm{PV}\) and then extend the consideration  to the case to the skyrmion lattices. The antivortices compensating the magnetic flux of the \(\mathrm{PVs}\) are located outside the \(\mathrm{Sk}\).  We analyzed the stable topological configurations and show that the vortex fields offer a way of manipulating the skyrmionic sizes.\\

\begin{figure}[t]
	\centering
	\includegraphics[width=0.8\linewidth]{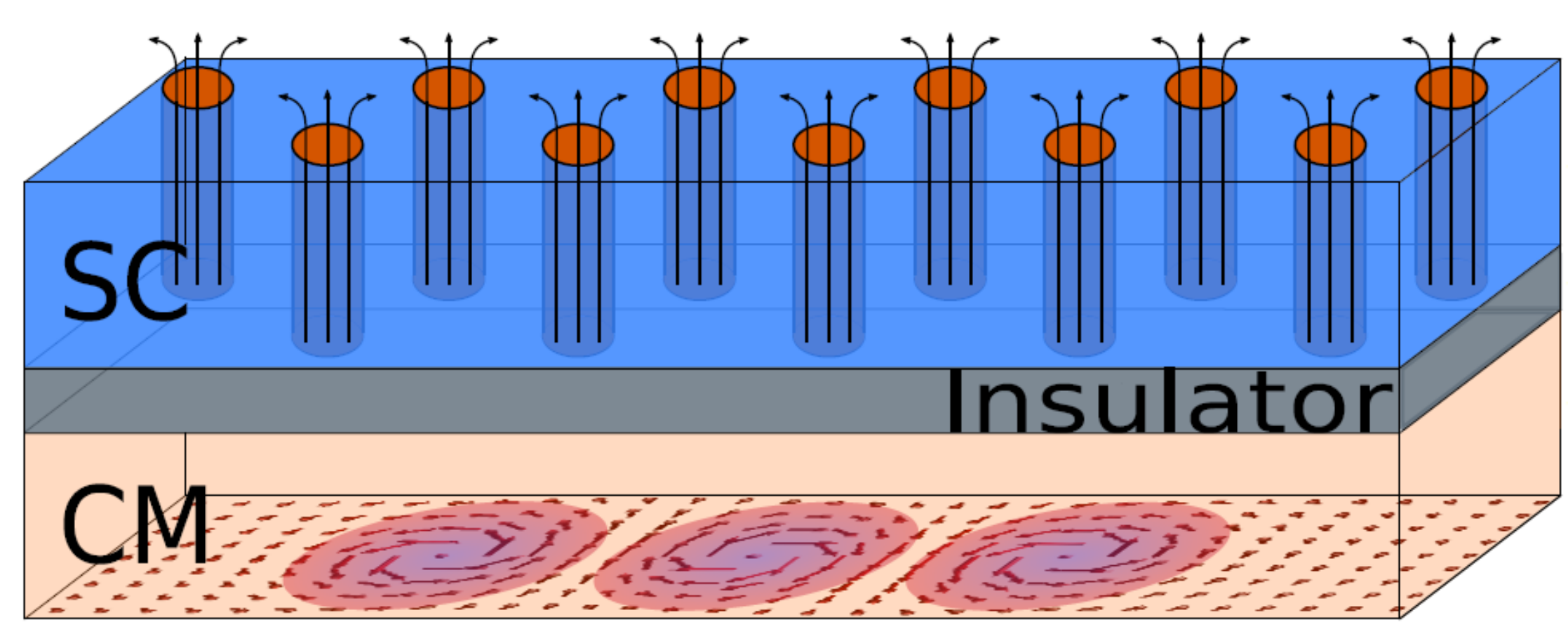}
	\caption{The heterostructure considered in the present paper. A chiral ferromagnet \(\mathrm{CM}\) with a skyrmion crystal \((\mathrm{SkX})\) is interfaced to a thin superconducting film \(\mathrm{SC}\) via a thin insulating barrier such that the interaction occurs only via the magnetic fields generated by the magnetic skyrmion. Note that the vortices in the superconducting film will be generated by the attractive vortex-skyrmion interaction.}
	\label{fig1}
\end{figure}

\textit{Model}: We consider a heterostructure composed by a chiral ferromagnet with magnetic skyrmion and a thin superconducting film interfacing the ferromagnet as shown in Fig.\ref{fig1}. We neglect a possible proximity effect assuming that it is weak or absent (in the case of an insulating layer between the \(\mathrm{SC}\) and \(\mathrm{F}\) film). A typical heterostructure we have had in mind consists of a thin film of the chiral ferromagnet like FeGe or Fe$_{0.5}$Co$_{0.5}$Si where skyrmions has been observed in a broad temperature range down to low temperatures \cite{Yu2011,Park2014} and an elemental superconductor such as Nb  or Al prepared using the standard sputtering method, employed typically in the superconducting spintronics \cite{Khaydukov2015}. The only requirement to the insulating layer is that its thickness must be thinner than the thicknesses of the S and F layers. Furthermore, even in the absence of an insulating layer the S/F interface transparency is often very low due to a mismatch of parameters of S and F materials such as impurity concentration, Fermi velocity etc.
%
In particular, we assume that the coupling between chiral magnet and a superconductors occurs solely via magnetic fields generated by the \(\mathrm{PVs}\) and \(\mathrm{Sk}\).
In thin superconducting films Abrikosov vortices \cite{abrikosov_magnetic_1957,abrikosov_fundamentals_1988} are transformed into Pearl vortices (PV) which are characterized by a weakly screened vortex field such that the field mainly runs outside the superconductor. As was shown previously, these can be spontaneously generated  in superconducting/ferromagnetic (SF) bilayers\cite{lyuksyutov_magnetization_1998,erdin_topological_2001,Milosevic2003,Lyuksyutov,Laiho2003,Burmistrov2005,Vestgarden2007}.  In order to find the equilibrium structure of the magnetization \(\vb{M}\) inside the ferromagnet and the location of the \(\mathrm{PVs}\) and \(\overline{\mathrm{PV}}s\) in the superconducting film, we have to find a minimum of the total free energy of the system \(E_{\text{tot}}\) which can be written in the following form 
	\begin{equation}
		E_{\text{tot}}=E_{\text{mag}}+E_{\text{Sk},\text{V}}+E_{\text{Sk},\overline{\text{V}}}+E_{\text{V},\text{V}}+E_{\text{V},\overline{\text{V}}}+E_{\overline{\text{V}},\overline{\text{V}}}\label{FreeEnergy}
	\end{equation}
where \(E_{\text{tot}}\) consists of the energies associated with the \(\mathrm{Sk}\) and the \(\mathrm{PV}s\). The energies \(E_{\text{Sk},\text{V}}\) and \(E_{\text{Sk},\overline{\text{V}}}\) describe the interaction energy between a skyrmion and a \(\mathrm{PV}\) and \(\overline{\mathrm{PV}}\), respectively. Furthermore, the vortex-vortex, vortex-antivortex, and antivortex-antivortex interaction energies are given by \( E_{\text{V},\text{V}}\) \(E_{\overline{\text{V}},\overline{\text{V}}}\) and \(E_{\text{V},\overline{\text{V}}}\), respectively.
The location of vortices depends on the size and helicity of the skyrmions, and whether one considers an isolated \(\mathrm{Sk}\) or a skyrmion lattice and several other parameters of the system. 

The magnetic energy \(E_{\text{mag}}\) of chiral ferromagnet can be written in the form \cite{bogdanov1989thermodynamically,kawaguchi_skyrmionic_2016}
	\begin{equation}
		E_{\text{mag}}=\int\dd[3]{\rho}M_{0}\qty[J\qty(\nabla\vb{m})^{2}-D\vb{m}\cdot\qty(\nabla\times\vb{m})+\vb{H}_{\text{ext}}\cdot\vb{m}]\label{MagEnergy}
	\end{equation}
where \(J\) and \(D\) are the strength of the spin-exchange (exchange stifness) and Dzyaloshinskii-Moriya interactions. For simplicity we assume that \(M_{0}\) is a constant, but the obtained result is unchanged if a softening of the absolute value of the magnetic moment, \(M_{0}\), is taken into account. In the latter case, the term $ \propto(\nabla M_0)^{2} $ should be added to the energy \(E_{\text{mag}}\) \cite{rosler_spontaneous_2006}. The last term in Eq.(\ref{MagEnergy}) denotes the interaction of the \(\mathrm{Sk}\) with an external magnetic field \(\vb{H}_{\text{ext}}\). We assume that the field \(\vb{H}_{\text{ext}}\) is absent since the role of the external field can be played by the fields \(\vb{h}_{\text{V}}\)  and \(\vb{h}_{\overline{\text{V}}}\) generated by the \(\mathrm{PV}\) and \(\overline{\mathrm{PV}}\), respectively.
In the simplest case, the magnetization vector \(\vb{M}(\vb*{r})\) in the chiral ferromagnet is oriented along the \(z\)-direction in the center of a cylindrically symmetric skyrmion and changes its direction like a Bloch (or N\'{e}el) domain wall with increasing distance \(\rho\) from the center. The skyrmion radius \( r_{\text{Sk}}\) can be identified as the distance at which the magnetization \(M_{z}\) changes its sign. In general the magnetization profile of a magnetic skyrmion in the \(\mathrm{F}\) thin film can be described by the magnetization vector \(\vb{M}(\vb*{r})=M_{0}\vb{m}(\rho,\varphi)\) where \(M_{0}\) is the saturation magnetization and \(\vb{m}(\rho,\varphi)\) is equal to \cite{kawaguchi_skyrmionic_2016}
\begin{equation}
\vb{m}(\rho,\theta)=\sin(\theta(\rho))\qty[\cos(\psi)\hat{e}_{\rho}+\sin(\psi)\hat{e}_{\varphi}]+\cos(\theta(\rho))\hat{e}_{z}
\end{equation}
here \(\theta(\rho)\) describes the angular variation of the skyrmion with respect to the distance from the \(\mathrm{Sk}\) center and \(\psi\) is the helicity of the skyrmion. The helicity for a Bloch and N\'{e}el skyrmion is given by \(\psi=\frac{\pi}{2}\) and \(\psi=\pi\), respectively.\\

\textit{Single Vortex}: Consider first the case of a single \(\mathrm{PV}\) inside a \(\mathrm{Sk}\) at a distance \(R\) from the \(\mathrm{Sk}\) center of a single skyrmion assuming that the antivortex \(\overline{\mathrm{PV}}\) is located far away from the center of the \(\mathrm{Sk}\). We assume for a moment that such a vortex is generated spontaneously and later discuss the formal criteria for this to happen. For simplicity we consider a Bloch skyrmion.
\begin{figure}[t]
	\centering
	\includegraphics[trim=6 7 7 7,clip,width=1.0\linewidth]{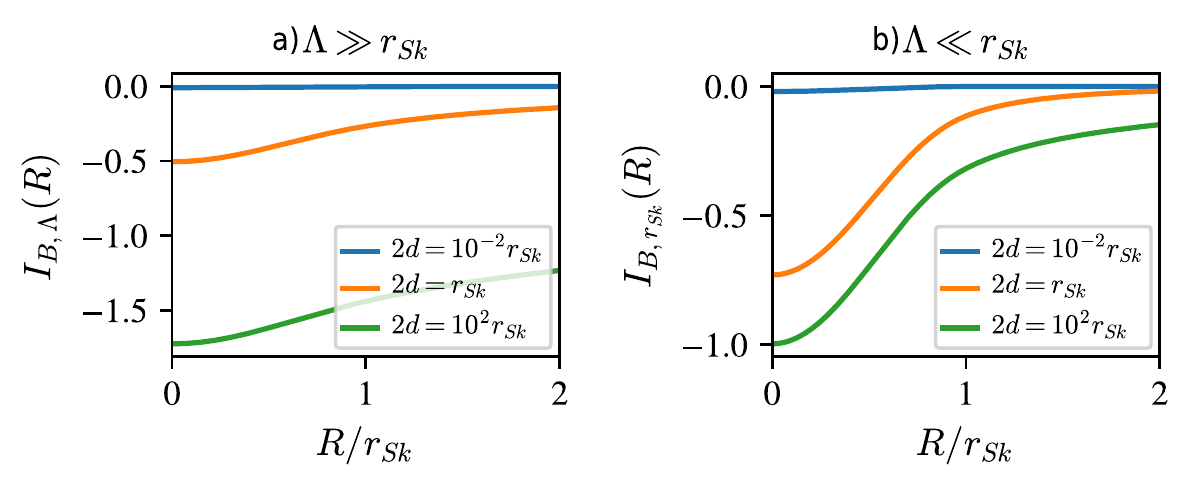}
	\vspace{-0.8cm}
	\caption{\label{ESKV} Dependence of the interaction energy between a Bloch \(\mathrm{Sk}\) and a \(\mathrm{PV}\) as a function of the distance \(R\) from the skyrmion center. \(I_{\text{B},\Lambda}\) (a) and \(I_{\text{B}, r_{\text{Sk}}}\) (b) are proportional to the energy  \(E_{\text{B},\text{V}}\) via Eq.(\ref{EBlochPV})}
\end{figure}  
Then the interaction energy between a \(\mathrm{PV}\) and the magnetization of a \(\mathrm{Sk}\) (Zeeman energy) is determined by the equation
	\begin{equation}
		E_{\text{Sk},\text{V}}=-M_{0}\int_{-d}^{d}\dd{z}\int\dd[2]{\rho}\vb{h}_{\text{V}}(\rho,z)\cdot\vb{m}
	\end{equation}
with \(2d\) being the thickness of the ferromagnet. The function \(\vb{h}_{\text{V}}\) describes the magnetic field created by the \(\mathrm{PV}\). It can be evaluated by solving London's equation, where the \(\mathrm{PV}\) enters the equation through a source term; the details are given in the Supplementary Material. In order to describe the interaction energy between a Bloch \(\mathrm{Sk}\) and a \(\mathrm{PV}\), \(E_{\text{B},\text{V}}\)=\(E_{\text{Sk},\text{V}}\), we only need to consider the coupling of the vortex-field with the \(z\)-component of the magnetization (since in the considered case of a Bloch \(\mathrm{Sk}\), \(m_{\rho}(k)=0\)), such that we obtain  
	\begin{eqnarray}
		\lefteqn{E_{\text{B},\text{V}}(R)  =  -M_{0}\phi_{0} r_{\text{Sk}} \times } && \nonumber \\
		&& \int\dd{k}2\sinh(k\tilde{d})e^{-k\tilde{d}}\frac{J_{0}\qty(kR/ r_{\text{Sk}})m_{z}(k)}{1+2k\Lambda/ r_{\text{Sk}}}\label{IntSk}
	\end{eqnarray}
with \(\tilde{d}=d/ r_{\text{Sk}}\) and the \(\mathrm{PV}\) size \(\Lambda=\lambda_{\text{L}}^{2}/2d_{\text{S}}\) where \(2d_{\text{S}}\) is the thickness of \(\mathrm{S}\) film and \(\lambda_{\text{L}}\) is the London penetration depth. Here \(k\) is a dimensionless quantity normalized w.r.t \(r_{\text{Sk}}\). The information about the spatial variation of the \(\mathrm{Sk}\) are contained in \(m_{z}(k)\) which is given by the Fourier transform of \(m_{z}(\rho,\theta)\), $m_{z}(k)=\int_{0}^{1}\dd{\rho'}\rho J_{0}\qty(k\rho')\qty[1+\cos(\theta(\rho'))]$
where \(J_{0}\) is the Bessel function of zeroth-order and with \(\rho'=\rho /r_{\text{Sk}}\).

Using a linear Ansatz for \mbox{\(\theta(\rho')=\pi\rho'\)} the magnetic energy \(E_{\text{mag}}\) can be simplified to
	\begin{equation}
		E_{\text{mag}}=2dM_{0}\qty[\frac{J a_{J}}{2}-D a_{D} r_{\text{Sk}}]
	\end{equation}
with the numerical coefficients \mbox{\(a_{J}=\pi(\pi^{2}+2.44)\)}, \mbox{\(a_{D}=\pi^{2}\)} \cite{kawaguchi_skyrmionic_2016}. To further simplify the expressions we consider two limiting cases (i) the size of the \(\mathrm{PV}\) is larger the size of the \(\mathrm{Sk}\) i.e. \(\Lambda\gg r_{\text{Sk}}\) or (ii) the opposite case, \(\Lambda\ll r_{\text{Sk}}\). The interaction energy  between the Bloch skyrmion and the Pearl vortex then acquires the following form
	\begin{equation}
		E_{\text{B},\text{V}}(R, r_{\text{Sk}})=M_{0}\phi_{0}\begin{cases}
		I_{\text{B},\Lambda}(R, r_{\text{Sk}}) r_{\text{Sk}}^{2}/2\Lambda,  \, &\text{for } \Lambda\gg r_{\text{Sk}}\\
		 I_{\text{B}, r_{\text{Sk}}}(R, r_{\text{Sk}})r_{\text{Sk}},  \, &\text{for } \Lambda\ll r_{\text{Sk}}
		\end{cases} \label{EBlochPV}
	\end{equation}
Here the functions \mbox{\(I_{\text{B},\Lambda}(R, r_{\text{Sk}})=-\int_{0}^{\infty}\dd{k}k^{-1}I(R, r_{\text{Sk}})\)} and \mbox{\(I_{\text{B}, r_{\text{Sk}}}(R, r_{\text{Sk}})=-\int_{0}^{\infty}\dd{k}I(R, r_{\text{Sk}})\)} where \mbox{\(I(R, r_{\text{Sk}})=2\sinh(k\tilde{d})e^{-k\tilde{d}}J_{0}(kR)m_{z}(k)\)}, describe the interaction of a \(\mathrm{PV}\) located at distance \(R\) from the center of the Bloch skyrmion. They are plotted in Fig. \ref{ESKV}. Observe that for $2d$ larger or of the same order as $ r_{\text{Sk}}$ the interaction between Pearl vortex and Bloch skyrmion is attractive (negative) in both cases for $R< r_{\text{Sk}}$, which is one of the most important results of our study. This indicates that once the Pearl vortex is created,  the interaction between \(\mathrm{Sk}\) and \(\mathrm{PV}s\) will guarantee the formation of the stable topological object.  Although the interaction between vortices can be neglected, to make sure the total flux in the superconductor is zero we still need to take into account the energy contribution \(2E_{\text{V}}\) needed for the creation of the \(\mathrm{PV}\)-\(\overline{\mathrm{PV}}\) pair, so that the total energy can be written as
	\begin{equation}
		E_{\text{B},\text{tot}}=E_{\text{mag}}+E_{\text{B},V}(R, r_{\text{Sk}})+\frac{\phi_{0}^{2}}{4\pi^{2}\Lambda}\ln(\frac{\Lambda}{ \xi})\label{SelfE}
	\end{equation}
where \(\xi\) (by assumption \(\xi\ll\Lambda\)) is the superconducting coherence length.
The last term is the doubled energy of a Pearl vortex \(E_{\text{V}}\) in the absence of an external magnetic field \cite{pearl_current_1964,gennes_superconductivity_1999,Lyuksyutov}. The stray field of the \(\mathrm{Sk}\) decreases this energy, but we neglect this correction which is small for a thin \(\mathrm{SC}\) film. Note that for a \(\mathrm{PV}\) to occur the condition \(\Lambda\gg \xi\) must be satisfied. If the vortex-vortex interaction is neglected, the \(\mathrm{PV}\) position is given by the  minima of the interaction energy \(E_{\text{Sk},\text{V}}\). For a Bloch \(\mathrm{Sk}\) this corresponds to the position at the skyrmion center \(R=0\).\\

{\it Skyrmion lattice and energy threshold for vortex-antivortex generation:} Following Ref.\cite{kawaguchi_skyrmionic_2016}, the energy of a skyrmion crystal (\(\mathrm{SkX}\)) can be obtained by multiplying \(E_{\text{SkX}}\) with the total number of \(\mathrm{Sks}\) fitting in our system of area \(L^{2}\). With this Ansatz  \(N=\frac{L}{\pi r_{\text{Sk}}^{2}}\) such that the total energy has now the form
	\begin{equation}
		E_{\text{SkX}} = \gamma\qty[\frac{J a_{J}}{2}-Da_{D} r_{\text{Sk}}+\frac{E_{\text{B,V}}(0, r_{\text{Sk}})}{2dM_{0}}+\frac{2E_{\text{V}}}{2dM_{0}}]
	\end{equation}
with \(\gamma=2dM_{0}L^{2}/\pi r_{\text{Sk}}^{2}\) and where we set \(R=0\) as it minimizes the energy independently from the size of the \(\mathrm{Sk}\).
In order to determine an analytic expression for the \(\mathrm{Sk}\) radius \( r_{\text{Sk}}\), we will consider a thin ferromagnetic film \(\tilde{d}\ll1\). Doing so we can factorize the \( r_{\text{Sk}}\) dependence out of \(I_{\text{B},\Lambda}\) and \(I_{\text{B}, r_{\text{Sk}}}\) resulting in  \(I_{\text{B},\Lambda}(0, r_{\text{Sk}})\approx-2d r_{\text{Sk}}^{-1} c_{1}\) and \(I_{\text{B}, r_{\text{Sk}}}(0, r_{\text{Sk}})\approx-2d r_{\text{Sk}}^{-1} c_{2}\) where \(c_{1}\approx0.86\) and \(c_{2}\approx2.13\). The total energy in the limit \(\Lambda\gg r_{\text{Sk}}\) can now be written as
	\begin{equation}
		E_{SkX}=\frac{2dM_{0}L^{2}}{\pi}\qty[\frac{\tilde{J}a_{J}}{2 r_{\text{Sk}}^{2}}-\frac{\tilde{D}a_{D}}{ r_{\text{Sk}}}],\quad \Lambda\ll r_{\text{Sk}}
	\end{equation}
For this limit the appearance of \(\mathrm{PV}\)-\(\overline{\mathrm{PV}}\) pairs leads to the renormalization of the interactions, \(\tilde{J}=J\qty[1+\frac{\phi_{0}^{2}\ln(\Lambda/ \xi)}{2\pi^{2}\Lambda2dM_{0}Ja_{J}}]\) and \(\tilde{D}=D\qty[1+\frac{\phi_{0}c_{1}}{Da_{D}2\Lambda}]\). In the case under consideration, the \(\mathrm{Sk}\) radius \(\tilde{r}_{\text{Sk}}\) is given by
\(\tilde{r}_{\text{Sk}}=a_{J}\tilde{J}/a_{D}\tilde{D}\).

The renormalization of the coefficients \(J\) and \(D\) results in an increased effective exchange stiffness \(\tilde{J}\) and increased effective  Dzyaloshinskii-Moriya constant \(\tilde{D}\). One can easily see that the presence of \(\mathrm{PV}s\) changes the ratio of the exchange and DMI constant resulting in an increase or decrease of the \(\mathrm{Sk}\) size \(\tilde{r}_{\text{Sk}}\). In particular, the \(\mathrm{Sk}\) shrinks if the condition 

\begin{equation}
	2dM_{0} r_{0}\geq \phi_{0}\ln(\frac{\Lambda}{\xi})/\pi^{2}c_{1}\label{Comp1}
\end{equation}

is satisfied, where \(r_{0}\) is the \(\mathrm{Sk}\) radius in the absence of vortices. Most importantly, we can now come back to the question on the energy threshold for the spontaneous  \(\mathrm{PV}\)-\(\overline{\mathrm{PV}}\) pair creation and show they indeed lower the total energy of our system. In particular, the difference of the energy \(\delta E = E_{\text{tot,B}}(0,\tilde{r
}_{\text{Sk}})-E_{\text{mag}}( r_{0})\) is equal to
	\begin{equation}
		\delta E = -\frac{2dM_{0}L^{2}}{\pi}\frac{a_{D}^{2}}{2a_{J}}\qty[\frac{\tilde{D}^{2}}{\tilde{J}}-\frac{D^{2}}{J}],\quad \Lambda\gg r_{\text{Sk}}\label{EnergyCompare1}
	\end{equation}
Eq.(\ref{EnergyCompare1}) shows that the energy of the system in the presence of vortices is lowered if the condition (\ref{Comp1}) is satisfied.

In the opposite limit \(\Lambda\ll r_{\text{Sk}}\) the total energy of a \(\mathrm{SkX}\) has the following form
	\begin{equation}
		E_{\text{SkX}}=\frac{2dM_{0}L^{2}}{\pi}\qty[\frac{\tilde{J}a_{J}}{2 r_{\text{Sk}}^{2}}-\frac{Da_{D}}{ r_{\text{Sk}}}] ,\quad \Lambda\ll r_{\text{Sk}}
	\end{equation}
The contributions of the \(\mathrm{PV}\) energy \(2E_{\text{V}}\) and the \(\mathrm{Sk}\)-\(\mathrm{PV}\) interaction energy \(E_{\text{B},\text{V}}\)  renormalize the exchange stiffness \(J\)  to $\tilde{J}=J\qty[1-\frac{2\phi_{0}c_{2}}{Ja_{J}}+\frac{\phi_{0}^{2}\ln(\Lambda/ \xi)}{2\pi^{2}\Lambda 2dM_{0}Ja_{J}}]$, whereas \(D\) remains the same.
Again, we can easily show that the energy of the system is lowered and the \(\mathrm{Sk}\) radius shrinks if the condition
\begin{equation}
	2dM_{0} \Lambda\geq \phi_{0}\ln(\frac{\Lambda}{\xi})/\pi^{2}c_{2}\label{Comp2}
\end{equation}
is fulfilled. Thus, Eqs. (\ref{Comp1},\ref{Comp2}) determine a threshold value of the magnetic moment \(M_{0}\) for the creation of \(\mathrm{PV}\)-\(\overline{\mathrm{PV}}\) pairs in the two limiting cases: \(\Lambda\gg r_{\text{Sk}}\) and \(\Lambda\ll r_{\text{Sk}}\). We also evaluate this threshold numerically and depict it in Fig. 4(b).
%
\begin{figure}[t]
	\centering
	\includegraphics[width=0.7\linewidth]{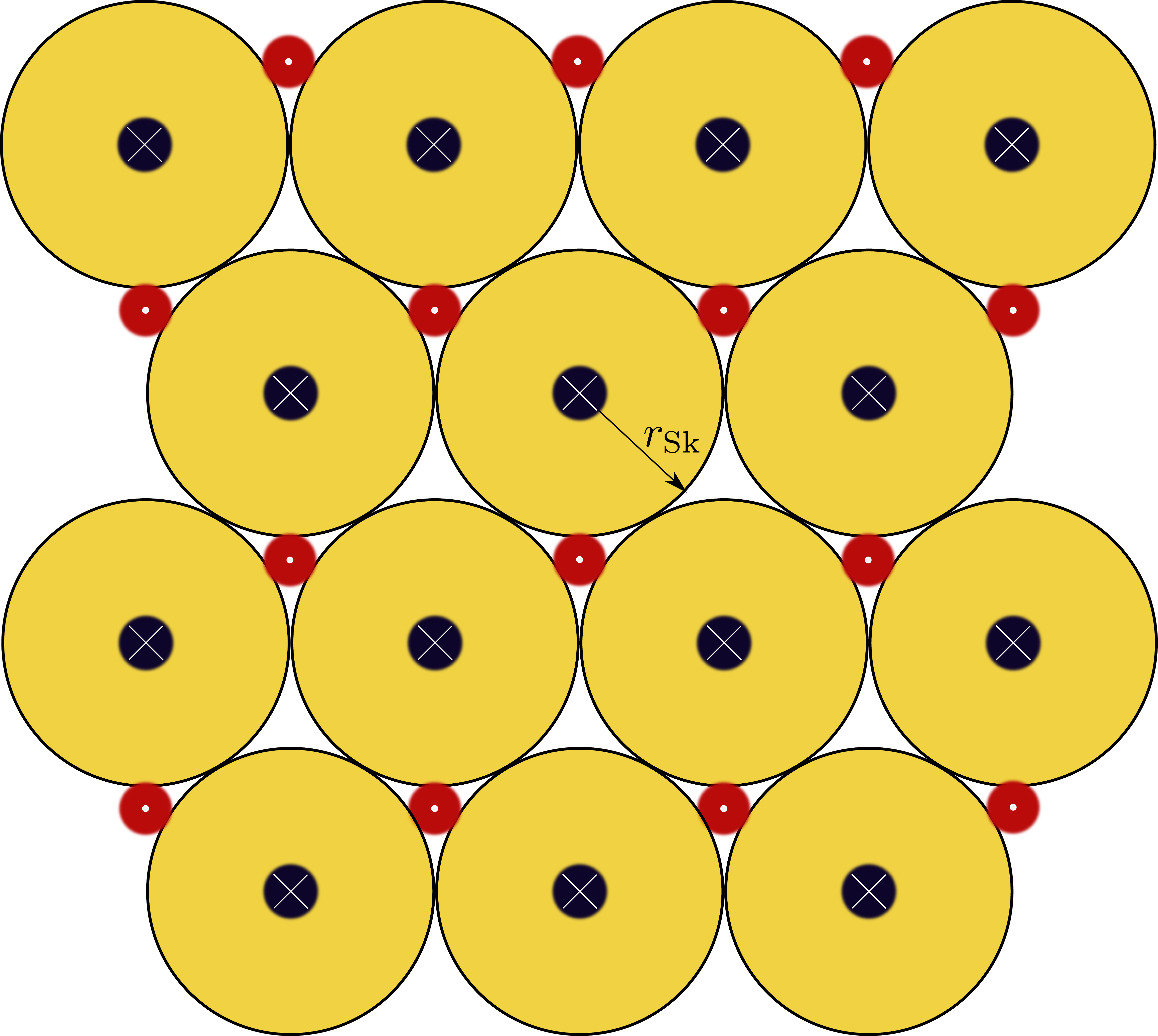}
	\caption{Schematic representation of the \(\mathrm{SkX}\) phase in the presence of the Abrikosov vortex lattice \(\mathrm{PV}\). The Anti-vortices(blue circles/white cross) are positioned above(or below) the vertices of the skyrmion unit cell. The \(\mathrm{PV}\) (red circles/white dot) are located above(or below) the \(\mathrm{Sks}\) center.\label{Schema}} 
\end{figure}	

\begin{figure}[t]
	\centering
	\includegraphics[trim=6 7 7 7,clip,width=1.0\linewidth]{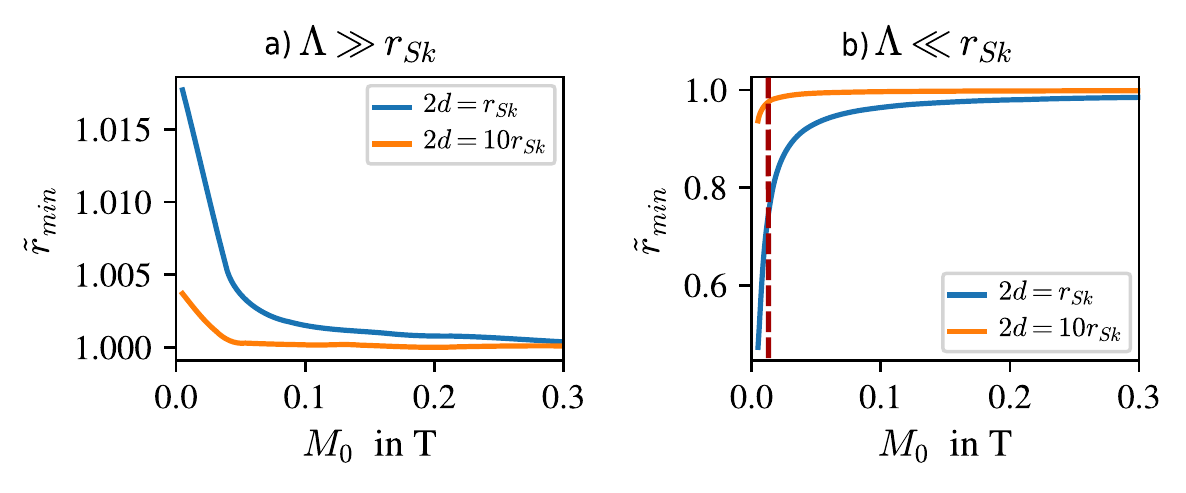}
	\vspace{-0.7cm}
	\caption{Plot of the Skyrmion radius in the presence of \(\mathrm{PV}s\) \(\tilde{r}_{min}=r_{\text{Sk}}/r_{0}\)  against saturation magnetization \(M_{0}\) for a) \(\Lambda\gg r_{\text{Sk}}\) and b) \(\Lambda\ll r_{\text{Sk}}\). The results were obtained for different ferromagnet thickness \(d\) and \(\xi=2.45\,\mathrm{nm}\), \(\Lambda=2450\,\mathrm{nm}\), \(J=10^{-7}\mathrm{erg}/\mathrm{cm}\) and \(D=0.08\,\mathrm{erg}/\mathrm{cm}^{2}\). Only b) is energetically favourable. The red dotted line in b) indicates a lower limit of \(M_{0}\), below which the \(\mathrm{PV}-\overline{\mathrm{PV}}\) creation is no longer energetically possible.\label{SkXr}}
\end{figure}

In a real system the periodic lattice configuration of the \(\mathrm{Sk}\)-lattice determines the position of Anti-Pearl vortices. The interaction between \(\overline{\mathrm{PV}}\) and \(\mathrm{Sk}\) is minimal at the vertices's of the \(Sk\)-lattice located at \(\tilde{R}=\sqrt{3}r_{\text{Sk}}\) from the \(\mathrm{Sk}\) center. Thus we can assume that the \(\bar{\mathrm{PV}}s\) are not positioned arbitrary far away but stay in the vicinity of the \(\mathrm{Sk}\) unit cell. By including the energy contributions associated with the \(\overline{\mathrm{PV}}s\) the free energy is given by
\begin{equation} 
E_{\text{SkX},\text{tot}}  =  E_{\text{SkX}}+  \gamma\qty[E_{\text{B},\bar{\text{V}}}(\sqrt{3}r_{\text{Sk}},r_{\text{Sk}})+E_{\text{V},\bar{V}}+E_{\bar{\text{V}},\bar{\text{V}}}]
\end{equation}
where \(E_{\text{B},\overline{\text{V}}}\) denotes the \(\overline{\mathrm{PV}}\)-\(\mathrm{Sk}\) interaction. The energy contributions \(E_{\text{V},\bar{V}}\) and \(E_{\bar{\text{V}},\bar{\text{V}}}\) associated with interactions of \(\overline{\mathrm{PV}}\) with \(\mathrm{PV}\) and other \(\overline{\mathrm{PV}}\) can be found in the appendix for the hybrid structure shown in  Fig.\ref{Schema}.
We performed a minimization of the total free energy, Eq. (14), with respect to the skyrmion size \(r_{\mathrm{Sk}}\). The result is shown in Fig.\ref{SkXr} for the two limiting cases \(\Lambda\gg r_{\text{Sk}}\) and \(\Lambda\ll r_{\text{Sk}}\), where we plotted \(\tilde{r}=r_{\text{Sk}}/r_{0}\) against the saturation magnetization \(M_{0}\) with \(r_{0}\) being the skyrmion radius in absence of \(\mathrm{PV}s\). We can see that as soon as the interaction of the \(\mathrm{PV}s\) and \(\overline{\mathrm{PV}}s\) gets weaker, the size of the skyrmion is changed. 
Comparing the energy of the system with \(\mathrm{PV}\)-\(\overline{\mathrm{PV}}\) pairs \(E_{\text{SkX}}(r_{Sk})\) with the energy without vortices \(E_{\text{SkX}}(r_{0})\). One can show that the case \(\Lambda\gg r_{\text{Sk}}\) is not energetically favorable which is due to the overlapping vortex regions of the \(\overline{\mathrm{PV}}\). However for \(\Lambda\ll r_{\text{Sk}}\) the overlap only occurs if the the size of the \(\mathrm{Sk}\) shrinks since the vortex positions are given in terms of \(r_{\text{Sk}}\), resulting in a lower energy of the system compared to a system without vortices. In this case, a weak interaction between \(\mathrm{Sk}s \) and vortices makes it favorable to decrease the \(\mathrm{Sk}\) size to maximize the overlap between the \(\mathrm{PV}\) and the \(\overline{\mathrm{PV}}s\), while keeping the interaction among \(\overline{\mathrm{PV}}s\) minimal. 

%

{\it Conclusions.} We considered a superconductor/chiral ferromagnet heterostructure with magnetic skyrmions in the ferromagnet. We showed that Pearl vortices appear spontaneously in the thin superconducting film as it happens in superconducting/ferromagnetic bilayers with ferromagnetic domains in the ferromagnetic film\cite{lyuksyutov_magnetization_1998,laiho_penetration_2003}. These \(\mathrm{PVs}\) lower the energy of the system and shrink the \(\mathrm{Sk}\) radius \( r_{\text{Sk}}\). The number of the occurring \(\mathrm{PVs}\) may be larger than one. It is determined by the balance of the force of attraction to the center of the \(\mathrm{Sk}\) and the force of repulsion between \(\mathrm{PVs}\). Anti-vortices \(\overline{\mathrm{PVs}}\) compensating the magnetic flux of the \(\mathrm{PV}\) are situated outside the skyrmion. The magnetic field created by \(\mathrm{PVs}\) acts on the skyrmion as an external magnetic field so that a \(\mathrm{Sk}\) lattice may exist in the system under consideration even in the absence of an external magnetic field.  Finally, we argue that the size of the skyrmions will be greatly affected by the presence of the vortices offering another prospect of manipulating the skyrmionic size by the proximity to a superconductor.

{\it Acknowledgements.} The authors acknowledge support from the DFG Priority Program SPP2137, “Skyrmionics,” under Grant No. ER 463/10.

\bibliography{literature} 

\begin{thebibliography}{46}%
\makeatletter
\providecommand \@ifxundefined [1]{%
 \@ifx{#1\undefined}
}%
\providecommand \@ifnum [1]{%
 \ifnum #1\expandafter \@firstoftwo
 \else \expandafter \@secondoftwo
 \fi
}%
\providecommand \@ifx [1]{%
 \ifx #1\expandafter \@firstoftwo
 \else \expandafter \@secondoftwo
 \fi
}%
\providecommand \natexlab [1]{#1}%
\providecommand \enquote  [1]{``#1''}%
\providecommand \bibnamefont  [1]{#1}%
\providecommand \bibfnamefont [1]{#1}%
\providecommand \citenamefont [1]{#1}%
\providecommand \href@noop [0]{\@secondoftwo}%
\providecommand \href [0]{\begingroup \@sanitize@url \@href}%
\providecommand \@href[1]{\@@startlink{#1}\@@href}%
\providecommand \@@href[1]{\endgroup#1\@@endlink}%
\providecommand \@sanitize@url [0]{\catcode `\\12\catcode `\$12\catcode
  `\&12\catcode `\#12\catcode `\^12\catcode `\_12\catcode `\%12\relax}%
\providecommand \@@startlink[1]{}%
\providecommand \@@endlink[0]{}%
\providecommand \url  [0]{\begingroup\@sanitize@url \@url }%
\providecommand \@url [1]{\endgroup\@href {#1}{\urlprefix }}%
\providecommand \urlprefix  [0]{URL }%
\providecommand \Eprint [0]{\href }%
\providecommand \doibase [0]{http://dx.doi.org/}%
\providecommand \selectlanguage [0]{\@gobble}%
\providecommand \bibinfo  [0]{\@secondoftwo}%
\providecommand \bibfield  [0]{\@secondoftwo}%
\providecommand \translation [1]{[#1]}%
\providecommand \BibitemOpen [0]{}%
\providecommand \bibitemStop [0]{}%
\providecommand \bibitemNoStop [0]{.\EOS\space}%
\providecommand \EOS [0]{\spacefactor3000\relax}%
\providecommand \BibitemShut  [1]{\csname bibitem#1\endcsname}%
\let\auto@bib@innerbib\@empty
\bibitem [{\citenamefont {Skyrme}(1962)}]{skyrme_unified_1962}%
  \BibitemOpen
  \bibfield  {author} {\bibinfo {author} {\bibfnamefont {T.~H.~R.}\
  \bibnamefont {Skyrme}},\ }\href {\doibase 10.1016/0029-5582(62)90775-7}
  {\bibfield  {journal} {\bibinfo  {journal} {Nuclear Physics}\ }\textbf
  {\bibinfo {volume} {31}},\ \bibinfo {pages} {556} (\bibinfo {year}
  {1962})}\BibitemShut {NoStop}%
\bibitem [{\citenamefont {M{\"u}hlbauer}\ \emph {et~al.}(2009)\citenamefont
  {M{\"u}hlbauer}, \citenamefont {Binz}, \citenamefont {Jonietz}, \citenamefont
  {Pfleiderer}, \citenamefont {Rosch}, \citenamefont {Neubauer}, \citenamefont
  {Georgii},\ and\ \citenamefont {B{\"o}ni}}]{muehlbauer_skyrmion_2009}%
  \BibitemOpen
  \bibfield  {author} {\bibinfo {author} {\bibfnamefont {S.}~\bibnamefont
  {M{\"u}hlbauer}}, \bibinfo {author} {\bibfnamefont {B.}~\bibnamefont {Binz}},
  \bibinfo {author} {\bibfnamefont {F.}~\bibnamefont {Jonietz}}, \bibinfo
  {author} {\bibfnamefont {C.}~\bibnamefont {Pfleiderer}}, \bibinfo {author}
  {\bibfnamefont {A.}~\bibnamefont {Rosch}}, \bibinfo {author} {\bibfnamefont
  {A.}~\bibnamefont {Neubauer}}, \bibinfo {author} {\bibfnamefont
  {R.}~\bibnamefont {Georgii}}, \ and\ \bibinfo {author} {\bibfnamefont
  {P.}~\bibnamefont {B{\"o}ni}},\ }\href {\doibase 10.1126/science.1166767}
  {\bibfield  {journal} {\bibinfo  {journal} {Science}\ }\textbf {\bibinfo
  {volume} {323}},\ \bibinfo {pages} {915} (\bibinfo {year}
  {2009})}\BibitemShut {NoStop}%
\bibitem [{\citenamefont
  {Dzyaloshinsky}(1958)}]{dzyaloshinsky_thermodynamic_1958}%
  \BibitemOpen
  \bibfield  {author} {\bibinfo {author} {\bibfnamefont {I.}~\bibnamefont
  {Dzyaloshinsky}},\ }\href {\doibase
  https://doi.org/10.1016/0022-3697(58)90076-3} {\bibfield  {journal} {\bibinfo
   {journal} {J. Phys. Chem. Solids}\ }\textbf {\bibinfo {volume} {4}},\
  \bibinfo {pages} {241 } (\bibinfo {year} {1958})}\BibitemShut {NoStop}%
\bibitem [{\citenamefont {Moriya}(1960)}]{moriya_anisotropic_1960}%
  \BibitemOpen
  \bibfield  {author} {\bibinfo {author} {\bibfnamefont {T.}~\bibnamefont
  {Moriya}},\ }\href {\doibase 10.1103/PhysRev.120.91} {\bibfield  {journal}
  {\bibinfo  {journal} {Phys. Rev.}\ }\textbf {\bibinfo {volume} {120}},\
  \bibinfo {pages} {91} (\bibinfo {year} {1960})}\BibitemShut {NoStop}%
\bibitem [{\citenamefont {Bogdanov}\ and\ \citenamefont
  {Yablonskii}(1989)}]{bogdanov1989thermodynamically}%
  \BibitemOpen
  \bibfield  {author} {\bibinfo {author} {\bibfnamefont {A.~N.}\ \bibnamefont
  {Bogdanov}}\ and\ \bibinfo {author} {\bibfnamefont {D.}~\bibnamefont
  {Yablonskii}},\ }\href@noop {} {\bibfield  {journal} {\bibinfo  {journal}
  {Sov. Phys. JETP}\ }\textbf {\bibinfo {volume} {68}},\ \bibinfo {pages} {178}
  (\bibinfo {year} {1989})}\BibitemShut {NoStop}%
\bibitem [{\citenamefont {Yu}\ \emph {et~al.}(2010)\citenamefont {Yu},
  \citenamefont {Onose}, \citenamefont {Kanazawa}, \citenamefont {Park},
  \citenamefont {Han}, \citenamefont {Matsui}, \citenamefont {Nagaosa},\ and\
  \citenamefont {Tokura}}]{yu_real-space_2010}%
  \BibitemOpen
  \bibfield  {author} {\bibinfo {author} {\bibfnamefont {X.~Z.}\ \bibnamefont
  {Yu}}, \bibinfo {author} {\bibfnamefont {Y.}~\bibnamefont {Onose}}, \bibinfo
  {author} {\bibfnamefont {N.}~\bibnamefont {Kanazawa}}, \bibinfo {author}
  {\bibfnamefont {J.~H.}\ \bibnamefont {Park}}, \bibinfo {author}
  {\bibfnamefont {J.~H.}\ \bibnamefont {Han}}, \bibinfo {author} {\bibfnamefont
  {Y.}~\bibnamefont {Matsui}}, \bibinfo {author} {\bibfnamefont
  {N.}~\bibnamefont {Nagaosa}}, \ and\ \bibinfo {author} {\bibfnamefont
  {Y.}~\bibnamefont {Tokura}},\ }\href {http://dx.doi.org/10.1038/nature09124}
  {\bibfield  {journal} {\bibinfo  {journal} {Nature}\ }\textbf {\bibinfo
  {volume} {465}},\ \bibinfo {pages} {901} (\bibinfo {year}
  {2010})}\BibitemShut {NoStop}%
\bibitem [{\citenamefont {Heinze}\ \emph
  {et~al.}(2011{\natexlab{a}})\citenamefont {Heinze}, \citenamefont {von
  Bergmann}, \citenamefont {Menzel}, \citenamefont {Brede}, \citenamefont
  {Kubetzka}, \citenamefont {Wiesendanger}, \citenamefont {Bihlmayer},\ and\
  \citenamefont {Blügel}}]{heinze_spontaneous_2011}%
  \BibitemOpen
  \bibfield  {author} {\bibinfo {author} {\bibfnamefont {S.}~\bibnamefont
  {Heinze}}, \bibinfo {author} {\bibfnamefont {K.}~\bibnamefont {von
  Bergmann}}, \bibinfo {author} {\bibfnamefont {M.}~\bibnamefont {Menzel}},
  \bibinfo {author} {\bibfnamefont {J.}~\bibnamefont {Brede}}, \bibinfo
  {author} {\bibfnamefont {A.}~\bibnamefont {Kubetzka}}, \bibinfo {author}
  {\bibfnamefont {R.}~\bibnamefont {Wiesendanger}}, \bibinfo {author}
  {\bibfnamefont {G.}~\bibnamefont {Bihlmayer}}, \ and\ \bibinfo {author}
  {\bibfnamefont {S.}~\bibnamefont {Blügel}},\ }\href
  {http://dx.doi.org/10.1038/nphys2045} {\bibfield  {journal} {\bibinfo
  {journal} {Nature Physics}\ }\textbf {\bibinfo {volume} {7}},\ \bibinfo
  {pages} {713} (\bibinfo {year} {2011}{\natexlab{a}})}\BibitemShut {NoStop}%
\bibitem [{\citenamefont {Fert}\ \emph {et~al.}(2017)\citenamefont {Fert},
  \citenamefont {Reyren},\ and\ \citenamefont {Cros}}]{fert_magnetic_2017}%
  \BibitemOpen
  \bibfield  {author} {\bibinfo {author} {\bibfnamefont {A.}~\bibnamefont
  {Fert}}, \bibinfo {author} {\bibfnamefont {N.}~\bibnamefont {Reyren}}, \ and\
  \bibinfo {author} {\bibfnamefont {V.}~\bibnamefont {Cros}},\ }\href
  {http://dx.doi.org/10.1038/natrevmats.2017.31} {\bibfield  {journal}
  {\bibinfo  {journal} {Nature Reviews Materials}\ }\textbf {\bibinfo {volume}
  {2}},\ \bibinfo {pages} {17031} (\bibinfo {year} {2017})}\BibitemShut
  {NoStop}%
\bibitem [{\citenamefont {Heinze}\ \emph
  {et~al.}(2011{\natexlab{b}})\citenamefont {Heinze}, \citenamefont {von
  Bergmann}, \citenamefont {Menzel}, \citenamefont {Brede}, \citenamefont
  {Kubetzka}, \citenamefont {Wiesendanger}, \citenamefont {Bihlmayer},\ and\
  \citenamefont {Bl\"ugel}}]{Heinze_2011}%
  \BibitemOpen
  \bibfield  {author} {\bibinfo {author} {\bibfnamefont {S.}~\bibnamefont
  {Heinze}}, \bibinfo {author} {\bibfnamefont {K.}~\bibnamefont {von
  Bergmann}}, \bibinfo {author} {\bibfnamefont {M.}~\bibnamefont {Menzel}},
  \bibinfo {author} {\bibfnamefont {J.}~\bibnamefont {Brede}}, \bibinfo
  {author} {\bibfnamefont {A.}~\bibnamefont {Kubetzka}}, \bibinfo {author}
  {\bibfnamefont {R.}~\bibnamefont {Wiesendanger}}, \bibinfo {author}
  {\bibfnamefont {G.}~\bibnamefont {Bihlmayer}}, \ and\ \bibinfo {author}
  {\bibfnamefont {S.}~\bibnamefont {Bl\"ugel}},\ }\href {\doibase
  10.1038/nphys2045} {\bibfield  {journal} {\bibinfo  {journal} {Nat. Phys.}\
  }\textbf {\bibinfo {volume} {7}},\ \bibinfo {pages} {713} (\bibinfo {year}
  {2011}{\natexlab{b}})}\BibitemShut {NoStop}%
\bibitem [{\citenamefont {Bogdanov}\ and\ \citenamefont
  {Hubert}(1994)}]{bogdanov_thermodynamically_1994}%
  \BibitemOpen
  \bibfield  {author} {\bibinfo {author} {\bibfnamefont {A.}~\bibnamefont
  {Bogdanov}}\ and\ \bibinfo {author} {\bibfnamefont {A.}~\bibnamefont
  {Hubert}},\ }\href {\doibase https://doi.org/10.1016/0304-8853(94)90046-9}
  {\bibfield  {journal} {\bibinfo  {journal} {J. Magn. Magn. Mater.}\ }\textbf
  {\bibinfo {volume} {138}},\ \bibinfo {pages} {255 } (\bibinfo {year}
  {1994})}\BibitemShut {NoStop}%
\bibitem [{\citenamefont {Bogdanov}(1995)}]{bogdanov1995new}%
  \BibitemOpen
  \bibfield  {author} {\bibinfo {author} {\bibfnamefont {A.}~\bibnamefont
  {Bogdanov}},\ }\href@noop {} {\bibfield  {journal} {\bibinfo  {journal} {JETP
  Letters}\ }\textbf {\bibinfo {volume} {62}},\ \bibinfo {pages} {247}
  (\bibinfo {year} {1995})}\BibitemShut {NoStop}%
\bibitem [{\citenamefont {R\"oßler}\ \emph {et~al.}(2006)\citenamefont
  {R\"oßler}, \citenamefont {Bogdanov},\ and\ \citenamefont
  {Pfleiderer}}]{rosler_spontaneous_2006}%
  \BibitemOpen
  \bibfield  {author} {\bibinfo {author} {\bibfnamefont {U.~K.}\ \bibnamefont
  {R\"oßler}}, \bibinfo {author} {\bibfnamefont {A.~N.}\ \bibnamefont
  {Bogdanov}}, \ and\ \bibinfo {author} {\bibfnamefont {C.}~\bibnamefont
  {Pfleiderer}},\ }\href {http://dx.doi.org/10.1038/nature05056} {\bibfield
  {journal} {\bibinfo  {journal} {Nature}\ }\textbf {\bibinfo {volume} {442}},\
  \bibinfo {pages} {797} (\bibinfo {year} {2006})}\BibitemShut {NoStop}%
\bibitem [{\citenamefont {Leonov}\ \emph {et~al.}(2016)\citenamefont {Leonov},
  \citenamefont {Monchesky}, \citenamefont {Romming}, \citenamefont {Kubetzka},
  \citenamefont {Bogdanov},\ and\ \citenamefont {Wiesendanger}}]{leonov2016}%
  \BibitemOpen
  \bibfield  {author} {\bibinfo {author} {\bibfnamefont {A.}~\bibnamefont
  {Leonov}}, \bibinfo {author} {\bibfnamefont {T.}~\bibnamefont {Monchesky}},
  \bibinfo {author} {\bibfnamefont {N.}~\bibnamefont {Romming}}, \bibinfo
  {author} {\bibfnamefont {A.}~\bibnamefont {Kubetzka}}, \bibinfo {author}
  {\bibfnamefont {A.}~\bibnamefont {Bogdanov}}, \ and\ \bibinfo {author}
  {\bibfnamefont {R.}~\bibnamefont {Wiesendanger}},\ }\href {\doibase
  10.1088/1367-2630/18/6/065003} {\bibfield  {journal} {\bibinfo  {journal}
  {New Journal of Physics}\ }\textbf {\bibinfo {volume} {18}},\ \bibinfo
  {pages} {065003} (\bibinfo {year} {2016})}\BibitemShut {NoStop}%
\bibitem [{\citenamefont {Bergeret}\ \emph {et~al.}(2005)\citenamefont
  {Bergeret}, \citenamefont {Volkov},\ and\ \citenamefont {Efetov}}]{Ber_2005}%
  \BibitemOpen
  \bibfield  {author} {\bibinfo {author} {\bibfnamefont {F.~S.}\ \bibnamefont
  {Bergeret}}, \bibinfo {author} {\bibfnamefont {A.~F.}\ \bibnamefont
  {Volkov}}, \ and\ \bibinfo {author} {\bibfnamefont {K.~B.}\ \bibnamefont
  {Efetov}},\ }\href {\doibase 10.1103/RevModPhys.77.1321} {\bibfield
  {journal} {\bibinfo  {journal} {Rev. Mod. Phys.}\ }\textbf {\bibinfo {volume}
  {77}},\ \bibinfo {pages} {1321} (\bibinfo {year} {2005})}\BibitemShut
  {NoStop}%
\bibitem [{\citenamefont {Buzdin}(2005)}]{Buz_2005}%
  \BibitemOpen
  \bibfield  {author} {\bibinfo {author} {\bibfnamefont {A.}~\bibnamefont
  {Buzdin}},\ }\href {\doibase 10.1103/RevModPhys.77.935} {\bibfield  {journal}
  {\bibinfo  {journal} {Rev. Mod. Phys.}\ }\textbf {\bibinfo {volume} {77}},\
  \bibinfo {pages} {935} (\bibinfo {year} {2005})}\BibitemShut {NoStop}%
\bibitem [{\citenamefont {Eschrig}(2011)}]{eschrig2011spin}%
  \BibitemOpen
  \bibfield  {author} {\bibinfo {author} {\bibfnamefont {M.}~\bibnamefont
  {Eschrig}},\ }\href {https://doi.org/10.1063/1.3541944} {\bibfield  {journal}
  {\bibinfo  {journal} {Phys. Today}\ }\textbf {\bibinfo {volume} {64}},\
  \bibinfo {pages} {43} (\bibinfo {year} {2011})}\BibitemShut {NoStop}%
\bibitem [{\citenamefont {Eschrig}(2015)}]{eschrig2015spin}%
  \BibitemOpen
  \bibfield  {author} {\bibinfo {author} {\bibfnamefont {M.}~\bibnamefont
  {Eschrig}},\ }\href {https://doi.org/10.1088/0034-4885/78/10/104501}
  {\bibfield  {journal} {\bibinfo  {journal} {Reports on Progress in Physics}\
  }\textbf {\bibinfo {volume} {78}},\ \bibinfo {pages} {104501} (\bibinfo
  {year} {2015})}\BibitemShut {NoStop}%
\bibitem [{\citenamefont {Linder}\ and\ \citenamefont
  {Balatsky}(2017)}]{balatsky_odd_2017}%
  \BibitemOpen
  \bibfield  {author} {\bibinfo {author} {\bibfnamefont {J.}~\bibnamefont
  {Linder}}\ and\ \bibinfo {author} {\bibfnamefont {A.}~\bibnamefont
  {Balatsky}},\ }\href {https://arxiv.org/abs/1709.03986} {\bibfield  {journal}
  {\bibinfo  {journal} {arXiv:1709.03986}\ } (\bibinfo {year}
  {2017})}\BibitemShut {NoStop}%
\bibitem [{\citenamefont {Linder}\ and\ \citenamefont
  {Robinson}(2015)}]{Lin_2015}%
  \BibitemOpen
  \bibfield  {author} {\bibinfo {author} {\bibfnamefont {J.}~\bibnamefont
  {Linder}}\ and\ \bibinfo {author} {\bibfnamefont {J.}~\bibnamefont
  {Robinson}},\ }\href {\doibase 10.1038/nphys3242} {\bibfield  {journal}
  {\bibinfo  {journal} {Nature Phys.}\ }\textbf {\bibinfo {volume} {11}},\
  \bibinfo {pages} {307} (\bibinfo {year} {2015})}\BibitemShut {NoStop}%
\bibitem [{\citenamefont {Sosnin}\ \emph {et~al.}(2006)\citenamefont {Sosnin},
  \citenamefont {Cho}, \citenamefont {Petrashov},\ and\ \citenamefont
  {Volkov}}]{Sos_2006}%
  \BibitemOpen
  \bibfield  {author} {\bibinfo {author} {\bibfnamefont {I.}~\bibnamefont
  {Sosnin}}, \bibinfo {author} {\bibfnamefont {H.}~\bibnamefont {Cho}},
  \bibinfo {author} {\bibfnamefont {V.~T.}\ \bibnamefont {Petrashov}}, \ and\
  \bibinfo {author} {\bibfnamefont {A.~F.}\ \bibnamefont {Volkov}},\ }\href
  {\doibase 10.1103/PhysRevLett.96.157002} {\bibfield  {journal} {\bibinfo
  {journal} {Phys. Rev. Lett.}\ }\textbf {\bibinfo {volume} {96}},\ \bibinfo
  {pages} {157002} (\bibinfo {year} {2006})}\BibitemShut {NoStop}%
\bibitem [{\citenamefont {Halasz}\ \emph {et~al.}(2009)\citenamefont {Halasz},
  \citenamefont {Robinson}, \citenamefont {Annett},\ and\ \citenamefont
  {Blamire}}]{Hal_2009}%
  \BibitemOpen
  \bibfield  {author} {\bibinfo {author} {\bibfnamefont {G.~B.}\ \bibnamefont
  {Halasz}}, \bibinfo {author} {\bibfnamefont {J.}~\bibnamefont {Robinson}},
  \bibinfo {author} {\bibfnamefont {J.~F.}\ \bibnamefont {Annett}}, \ and\
  \bibinfo {author} {\bibfnamefont {M.}~\bibnamefont {Blamire}},\ }\href
  {\doibase 10.1103/PhysRevB.79.224505} {\bibfield  {journal} {\bibinfo
  {journal} {Phys. Rev. B}\ }\textbf {\bibinfo {volume} {79}},\ \bibinfo
  {pages} {224505} (\bibinfo {year} {2009})}\BibitemShut {NoStop}%
\bibitem [{\citenamefont {Kalenkov}\ \emph {et~al.}(2011)\citenamefont
  {Kalenkov}, \citenamefont {Zaikin},\ and\ \citenamefont
  {Petrashov}}]{Kal_2011}%
  \BibitemOpen
  \bibfield  {author} {\bibinfo {author} {\bibfnamefont {M.~S.}\ \bibnamefont
  {Kalenkov}}, \bibinfo {author} {\bibfnamefont {A.~D.}\ \bibnamefont
  {Zaikin}}, \ and\ \bibinfo {author} {\bibfnamefont {V.~T.}\ \bibnamefont
  {Petrashov}},\ }\href {\doibase 10.1103/PhysRevLett.107.087003} {\bibfield
  {journal} {\bibinfo  {journal} {Phys. Rev. Lett.}\ }\textbf {\bibinfo
  {volume} {107}},\ \bibinfo {pages} {087003} (\bibinfo {year}
  {2011})}\BibitemShut {NoStop}%
\bibitem [{\citenamefont {Volkov}\ \emph {et~al.}(2003)\citenamefont {Volkov},
  \citenamefont {Bergeret},\ and\ \citenamefont {Efetov}}]{Volkov_2003}%
  \BibitemOpen
  \bibfield  {author} {\bibinfo {author} {\bibfnamefont {A.~F.}\ \bibnamefont
  {Volkov}}, \bibinfo {author} {\bibfnamefont {F.~S.}\ \bibnamefont
  {Bergeret}}, \ and\ \bibinfo {author} {\bibfnamefont {K.~B.}\ \bibnamefont
  {Efetov}},\ }\href {\doibase 10.1103/PhysRevLett.90.117006} {\bibfield
  {journal} {\bibinfo  {journal} {Phys. Rev. Lett.}\ }\textbf {\bibinfo
  {volume} {90}},\ \bibinfo {pages} {117006} (\bibinfo {year}
  {2003})}\BibitemShut {NoStop}%
\bibitem [{\citenamefont {Volkov}\ \emph {et~al.}(2006)\citenamefont {Volkov},
  \citenamefont {Anishchanka},\ and\ \citenamefont {Efetov}}]{Volkov_2006}%
  \BibitemOpen
  \bibfield  {author} {\bibinfo {author} {\bibfnamefont {A.~F.}\ \bibnamefont
  {Volkov}}, \bibinfo {author} {\bibfnamefont {A.}~\bibnamefont {Anishchanka}},
  \ and\ \bibinfo {author} {\bibfnamefont {K.~B.}\ \bibnamefont {Efetov}},\
  }\href {\doibase 10.1103/PhysRevB.73.104412} {\bibfield  {journal} {\bibinfo
  {journal} {Phys. Rev. B}\ }\textbf {\bibinfo {volume} {73}},\ \bibinfo
  {pages} {104412} (\bibinfo {year} {2006})}\BibitemShut {NoStop}%
\bibitem [{\citenamefont {Wu}\ \emph {et~al.}(2012)\citenamefont {Wu},
  \citenamefont {Valls},\ and\ \citenamefont {Halterman}}]{Wu2012}%
  \BibitemOpen
  \bibfield  {author} {\bibinfo {author} {\bibfnamefont {C.-T.}\ \bibnamefont
  {Wu}}, \bibinfo {author} {\bibfnamefont {O.~T.}\ \bibnamefont {Valls}}, \
  and\ \bibinfo {author} {\bibfnamefont {K.}~\bibnamefont {Halterman}},\ }\href
  {\doibase 10.1103/PhysRevB.86.184517} {\bibfield  {journal} {\bibinfo
  {journal} {Phys. Rev. B}\ }\textbf {\bibinfo {volume} {86}},\ \bibinfo
  {pages} {184517} (\bibinfo {year} {2012})}\BibitemShut {NoStop}%
\bibitem [{\citenamefont {Bjornson}\ and\ \citenamefont
  {Black-Schaffer}(2014)}]{Bjornson_2014}%
  \BibitemOpen
  \bibfield  {author} {\bibinfo {author} {\bibfnamefont {K.}~\bibnamefont
  {Bjornson}}\ and\ \bibinfo {author} {\bibfnamefont {A.~M.}\ \bibnamefont
  {Black-Schaffer}},\ }\href {\doibase 10.1103/PhysRevB.89.134518} {\bibfield
  {journal} {\bibinfo  {journal} {Phys. Rev. B}\ }\textbf {\bibinfo {volume}
  {89}},\ \bibinfo {pages} {134518} (\bibinfo {year} {2014})}\BibitemShut
  {NoStop}%
\bibitem [{\citenamefont {Yokoyama}\ and\ \citenamefont
  {Linder}(2015)}]{Yokoyama_2015}%
  \BibitemOpen
  \bibfield  {author} {\bibinfo {author} {\bibfnamefont {T.}~\bibnamefont
  {Yokoyama}}\ and\ \bibinfo {author} {\bibfnamefont {J.}~\bibnamefont
  {Linder}},\ }\href {\doibase 10.1103/PhysRevB.92.060503} {\bibfield
  {journal} {\bibinfo  {journal} {Phys. Rev. B}\ }\textbf {\bibinfo {volume}
  {92}},\ \bibinfo {pages} {060503(R)} (\bibinfo {year} {2015})}\BibitemShut
  {NoStop}%
\bibitem [{\citenamefont {Pershoguba}\ \emph {et~al.}(2016)\citenamefont
  {Pershoguba}, \citenamefont {Nakosai},\ and\ \citenamefont
  {Balatsky}}]{Pershoguba_2016}%
  \BibitemOpen
  \bibfield  {author} {\bibinfo {author} {\bibfnamefont {S.~S.}\ \bibnamefont
  {Pershoguba}}, \bibinfo {author} {\bibfnamefont {S.}~\bibnamefont {Nakosai}},
  \ and\ \bibinfo {author} {\bibfnamefont {A.~V.}\ \bibnamefont {Balatsky}},\
  }\href {\doibase 10.1103/PhysRevB.94.064513} {\bibfield  {journal} {\bibinfo
  {journal} {Phys. Rev. B}\ }\textbf {\bibinfo {volume} {94}},\ \bibinfo
  {pages} {064513} (\bibinfo {year} {2016})}\BibitemShut {NoStop}%
\bibitem [{\citenamefont {Bobkova}\ \emph {et~al.}(2018)\citenamefont
  {Bobkova}, \citenamefont {Bobkov},\ and\ \citenamefont
  {Silaev}}]{bobkova_spin_2018}%
  \BibitemOpen
  \bibfield  {author} {\bibinfo {author} {\bibfnamefont {I.~V.}\ \bibnamefont
  {Bobkova}}, \bibinfo {author} {\bibfnamefont {A.~M.}\ \bibnamefont {Bobkov}},
  \ and\ \bibinfo {author} {\bibfnamefont {M.~A.}\ \bibnamefont {Silaev}},\
  }\href {\doibase 10.1103/PhysRevB.98.014521} {\bibfield  {journal} {\bibinfo
  {journal} {Phys. Rev. B}\ }\textbf {\bibinfo {volume} {98}},\ \bibinfo
  {pages} {014521} (\bibinfo {year} {2018})}\BibitemShut {NoStop}%
\bibitem [{\citenamefont {Hals}\ \emph {et~al.}(2016)\citenamefont {Hals},
  \citenamefont {Schecter},\ and\ \citenamefont {Rudner}}]{Hals_2016}%
  \BibitemOpen
  \bibfield  {author} {\bibinfo {author} {\bibfnamefont {K.~M.}\ \bibnamefont
  {Hals}}, \bibinfo {author} {\bibfnamefont {M.}~\bibnamefont {Schecter}}, \
  and\ \bibinfo {author} {\bibfnamefont {M.~S.}\ \bibnamefont {Rudner}},\
  }\href {\doibase 10.1103/PhysRevLett.117.017001} {\bibfield  {journal}
  {\bibinfo  {journal} {Phys. Rev. Lett.}\ }\textbf {\bibinfo {volume} {117}},\
  \bibinfo {pages} {017001} (\bibinfo {year} {2016})}\BibitemShut {NoStop}%
\bibitem [{\citenamefont {Pearl}(1964)}]{pearl_current_1964}%
  \BibitemOpen
  \bibfield  {author} {\bibinfo {author} {\bibfnamefont {J.}~\bibnamefont
  {Pearl}},\ }\href {\doibase 10.1063/1.1754056} {\bibfield  {journal}
  {\bibinfo  {journal} {Appl. Phys. Lett.}\ }\textbf {\bibinfo {volume} {5}},\
  \bibinfo {pages} {65} (\bibinfo {year} {1964})}\BibitemShut {NoStop}%
\bibitem [{\citenamefont {G.~de Gennes}(1999)}]{gennes_superconductivity_1999}%
  \BibitemOpen
  \bibfield  {author} {\bibinfo {author} {\bibfnamefont {P.}~\bibnamefont
  {G.~de Gennes}},\ }\href@noop {} {\emph {\bibinfo {title} {Superconductivity
  of metals and alloys}}}\ (\bibinfo {year} {1999})\BibitemShut {NoStop}%
\bibitem [{\citenamefont {Yu}\ \emph {et~al.}(2011)\citenamefont {Yu},
  \citenamefont {Kanazawa}, \citenamefont {Onose}, \citenamefont {Kimoto},
  \citenamefont {Zhang}, \citenamefont {Ishiwata}, \citenamefont {Matsui},\
  and\ \citenamefont {Tokura}}]{Yu2011}%
  \BibitemOpen
  \bibfield  {author} {\bibinfo {author} {\bibfnamefont {X.~Z.}\ \bibnamefont
  {Yu}}, \bibinfo {author} {\bibfnamefont {N.}~\bibnamefont {Kanazawa}},
  \bibinfo {author} {\bibfnamefont {Y.}~\bibnamefont {Onose}}, \bibinfo
  {author} {\bibfnamefont {K.}~\bibnamefont {Kimoto}}, \bibinfo {author}
  {\bibfnamefont {W.~Z.}\ \bibnamefont {Zhang}}, \bibinfo {author}
  {\bibfnamefont {S.}~\bibnamefont {Ishiwata}}, \bibinfo {author}
  {\bibfnamefont {Y.}~\bibnamefont {Matsui}}, \ and\ \bibinfo {author}
  {\bibfnamefont {Y.}~\bibnamefont {Tokura}},\ }\href {\doibase
  10.1038/nmat2916} {\bibfield  {journal} {\bibinfo  {journal} {Nature
  Materials}\ }\textbf {\bibinfo {volume} {10}},\ \bibinfo {pages} {106}
  (\bibinfo {year} {2011})}\BibitemShut {NoStop}%
\bibitem [{\citenamefont {Park}\ \emph {et~al.}(2014)\citenamefont {Park},
  \citenamefont {Yu}, \citenamefont {Aizawa}, \citenamefont {Tanigaki},
  \citenamefont {Akashi}, \citenamefont {Takahashi}, \citenamefont {Matsuda},
  \citenamefont {Kanazawa}, \citenamefont {Onose}, \citenamefont {Shindo},
  \citenamefont {Tonomura},\ and\ \citenamefont {Tokura}}]{Park2014}%
  \BibitemOpen
  \bibfield  {author} {\bibinfo {author} {\bibfnamefont {H.~S.}\ \bibnamefont
  {Park}}, \bibinfo {author} {\bibfnamefont {X.}~\bibnamefont {Yu}}, \bibinfo
  {author} {\bibfnamefont {S.}~\bibnamefont {Aizawa}}, \bibinfo {author}
  {\bibfnamefont {T.}~\bibnamefont {Tanigaki}}, \bibinfo {author}
  {\bibfnamefont {T.}~\bibnamefont {Akashi}}, \bibinfo {author} {\bibfnamefont
  {Y.}~\bibnamefont {Takahashi}}, \bibinfo {author} {\bibfnamefont
  {T.}~\bibnamefont {Matsuda}}, \bibinfo {author} {\bibfnamefont
  {N.}~\bibnamefont {Kanazawa}}, \bibinfo {author} {\bibfnamefont
  {Y.}~\bibnamefont {Onose}}, \bibinfo {author} {\bibfnamefont
  {D.}~\bibnamefont {Shindo}}, \bibinfo {author} {\bibfnamefont
  {A.}~\bibnamefont {Tonomura}}, \ and\ \bibinfo {author} {\bibfnamefont
  {Y.}~\bibnamefont {Tokura}},\ }\href {\doibase 10.1038/nnano.2014.52}
  {\bibfield  {journal} {\bibinfo  {journal} {Nature Nanotechnology}\ }\textbf
  {\bibinfo {volume} {9}},\ \bibinfo {pages} {337} (\bibinfo {year}
  {2014})}\BibitemShut {NoStop}%
\bibitem [{\citenamefont {Khaydukov}\ \emph {et~al.}(2015)\citenamefont
  {Khaydukov}, \citenamefont {Morari}, \citenamefont {Soltwedel}, \citenamefont
  {Keller}, \citenamefont {Christiani}, \citenamefont {Logvenov}, \citenamefont
  {Kupriyanov}, \citenamefont {Sidorenko},\ and\ \citenamefont
  {Keimer}}]{Khaydukov2015}%
  \BibitemOpen
  \bibfield  {author} {\bibinfo {author} {\bibfnamefont {Y.}~\bibnamefont
  {Khaydukov}}, \bibinfo {author} {\bibfnamefont {R.}~\bibnamefont {Morari}},
  \bibinfo {author} {\bibfnamefont {O.}~\bibnamefont {Soltwedel}}, \bibinfo
  {author} {\bibfnamefont {T.}~\bibnamefont {Keller}}, \bibinfo {author}
  {\bibfnamefont {G.}~\bibnamefont {Christiani}}, \bibinfo {author}
  {\bibfnamefont {G.}~\bibnamefont {Logvenov}}, \bibinfo {author}
  {\bibfnamefont {M.}~\bibnamefont {Kupriyanov}}, \bibinfo {author}
  {\bibfnamefont {A.}~\bibnamefont {Sidorenko}}, \ and\ \bibinfo {author}
  {\bibfnamefont {B.}~\bibnamefont {Keimer}},\ }\href {\doibase
  10.1063/1.4936789} {\bibfield  {journal} {\bibinfo  {journal} {J. Appl.
  Phys.}\ }\textbf {\bibinfo {volume} {118}},\ \bibinfo {pages} {213905}
  (\bibinfo {year} {2015})}\BibitemShut {NoStop}%
\bibitem [{\citenamefont {Abrikosov}(1957)}]{abrikosov_magnetic_1957}%
  \BibitemOpen
  \bibfield  {author} {\bibinfo {author} {\bibfnamefont {A.}~\bibnamefont
  {Abrikosov}},\ }\href@noop {} {\bibfield  {journal} {\bibinfo  {journal}
  {Sov. Phys. JETP}\ }\textbf {\bibinfo {volume} {5}} (\bibinfo {year}
  {1957})}\BibitemShut {NoStop}%
\bibitem [{\citenamefont {Abrikosov}(1988)}]{abrikosov_fundamentals_1988}%
  \BibitemOpen
  \bibfield  {author} {\bibinfo {author} {\bibfnamefont {A.~A.}\ \bibnamefont
  {Abrikosov}},\ }\href@noop {} {\emph {\bibinfo {title} {Fundamentals of the
  theory of metals}}}\ (\bibinfo {year} {1988})\BibitemShut {NoStop}%
\bibitem [{\citenamefont {Lyuksyutov}\ and\ \citenamefont
  {Pokrovsky}(1998)}]{lyuksyutov_magnetization_1998}%
  \BibitemOpen
  \bibfield  {author} {\bibinfo {author} {\bibfnamefont {I.~F.}\ \bibnamefont
  {Lyuksyutov}}\ and\ \bibinfo {author} {\bibfnamefont {V.}~\bibnamefont
  {Pokrovsky}},\ }\href {\doibase 10.1103/PhysRevLett.81.2344} {\bibfield
  {journal} {\bibinfo  {journal} {Phys. Rev. Lett.}\ }\textbf {\bibinfo
  {volume} {81}},\ \bibinfo {pages} {2344} (\bibinfo {year}
  {1998})}\BibitemShut {NoStop}%
\bibitem [{\citenamefont {Erdin}\ \emph {et~al.}(2001)\citenamefont {Erdin},
  \citenamefont {Lyuksyutov}, \citenamefont {Pokrovsky},\ and\ \citenamefont
  {Vinokur}}]{erdin_topological_2001}%
  \BibitemOpen
  \bibfield  {author} {\bibinfo {author} {\bibfnamefont {S.}~\bibnamefont
  {Erdin}}, \bibinfo {author} {\bibfnamefont {I.~F.}\ \bibnamefont
  {Lyuksyutov}}, \bibinfo {author} {\bibfnamefont {V.~L.}\ \bibnamefont
  {Pokrovsky}}, \ and\ \bibinfo {author} {\bibfnamefont {V.~M.}\ \bibnamefont
  {Vinokur}},\ }\href {\doibase 10.1103/PhysRevLett.88.017001} {\bibfield
  {journal} {\bibinfo  {journal} {Phys. Rev. Lett.}\ }\textbf {\bibinfo
  {volume} {88}},\ \bibinfo {pages} {017001} (\bibinfo {year}
  {2001})}\BibitemShut {NoStop}%
\bibitem [{\citenamefont {Milosevic}\ and\ \citenamefont
  {Peeters}(2003)}]{Milosevic2003}%
  \BibitemOpen
  \bibfield  {author} {\bibinfo {author} {\bibfnamefont {M.}~\bibnamefont
  {Milosevic}}\ and\ \bibinfo {author} {\bibfnamefont {F.}~\bibnamefont
  {Peeters}},\ }\href {\doibase 10.1103/PhysRevB.68.094510} {\bibfield
  {journal} {\bibinfo  {journal} {Phys. Rev. B}\ }\textbf {\bibinfo {volume}
  {68}},\ \bibinfo {pages} {094510} (\bibinfo {year} {2003})}\BibitemShut
  {NoStop}%
\bibitem [{\citenamefont {Lyuksyutov}\ and\ \citenamefont
  {Pokrovsky}(2005)}]{Lyuksyutov}%
  \BibitemOpen
  \bibfield  {author} {\bibinfo {author} {\bibfnamefont {I.~F.}\ \bibnamefont
  {Lyuksyutov}}\ and\ \bibinfo {author} {\bibfnamefont {V.~L.}\ \bibnamefont
  {Pokrovsky}},\ }\href {\doibase 10.1080/00018730500057536} {\bibfield
  {journal} {\bibinfo  {journal} {Adv. Phys.}\ }\textbf {\bibinfo {volume}
  {54}},\ \bibinfo {pages} {67} (\bibinfo {year} {2005})}\BibitemShut {NoStop}%
\bibitem [{\citenamefont {Laiho}\ \emph
  {et~al.}(2003{\natexlab{a}})\citenamefont {Laiho}, \citenamefont
  {L\"ahderanta}, \citenamefont {Sonin},\ and\ \citenamefont
  {Traito}}]{Laiho2003}%
  \BibitemOpen
  \bibfield  {author} {\bibinfo {author} {\bibfnamefont {R.}~\bibnamefont
  {Laiho}}, \bibinfo {author} {\bibfnamefont {E.}~\bibnamefont {L\"ahderanta}},
  \bibinfo {author} {\bibfnamefont {E.}~\bibnamefont {Sonin}}, \ and\ \bibinfo
  {author} {\bibfnamefont {K.}~\bibnamefont {Traito}},\ }\href {\doibase
  10.1103/PhysRevB.67.144522} {\bibfield  {journal} {\bibinfo  {journal} {Phys.
  Rev. B}\ }\textbf {\bibinfo {volume} {67}},\ \bibinfo {pages} {144522}
  (\bibinfo {year} {2003}{\natexlab{a}})}\BibitemShut {NoStop}%
\bibitem [{\citenamefont {Burmistrov}\ and\ \citenamefont
  {Chtchelkatchev}(2005)}]{Burmistrov2005}%
  \BibitemOpen
  \bibfield  {author} {\bibinfo {author} {\bibfnamefont {I.}~\bibnamefont
  {Burmistrov}}\ and\ \bibinfo {author} {\bibfnamefont {N.}~\bibnamefont
  {Chtchelkatchev}},\ }\href {\doibase 10.1103/PhysRevB.72.144520} {\bibfield
  {journal} {\bibinfo  {journal} {Phys. Rev. B}\ }\textbf {\bibinfo {volume}
  {72}},\ \bibinfo {pages} {144520} (\bibinfo {year} {2005})}\BibitemShut
  {NoStop}%
\bibitem [{\citenamefont {Vestgarden}\ \emph {et~al.}(2007)\citenamefont
  {Vestgarden}, \citenamefont {Shantsev}, \citenamefont {Olsen}, \citenamefont
  {Galperin}, \citenamefont {Yurchenko}, \citenamefont {Goa},\ and\
  \citenamefont {Johansen}}]{Vestgarden2007}%
  \BibitemOpen
  \bibfield  {author} {\bibinfo {author} {\bibfnamefont {J.}~\bibnamefont
  {Vestgarden}}, \bibinfo {author} {\bibfnamefont {D.}~\bibnamefont
  {Shantsev}}, \bibinfo {author} {\bibfnamefont {A.}~\bibnamefont {Olsen}},
  \bibinfo {author} {\bibfnamefont {Y.}~\bibnamefont {Galperin}}, \bibinfo
  {author} {\bibfnamefont {V.}~\bibnamefont {Yurchenko}}, \bibinfo {author}
  {\bibfnamefont {P.}~\bibnamefont {Goa}}, \ and\ \bibinfo {author}
  {\bibfnamefont {T.}~\bibnamefont {Johansen}},\ }\href {\doibase
  10.1103/PhysRevLett.98.117202} {\bibfield  {journal} {\bibinfo  {journal}
  {Phys. Rev. Lett.}\ }\textbf {\bibinfo {volume} {98}},\ \bibinfo {pages}
  {117202} (\bibinfo {year} {2007})}\BibitemShut {NoStop}%
\bibitem [{\citenamefont {Kawaguchi}\ \emph {et~al.}(2016)\citenamefont
  {Kawaguchi}, \citenamefont {Tanaka},\ and\ \citenamefont
  {Nagaosa}}]{kawaguchi_skyrmionic_2016}%
  \BibitemOpen
  \bibfield  {author} {\bibinfo {author} {\bibfnamefont {Y.}~\bibnamefont
  {Kawaguchi}}, \bibinfo {author} {\bibfnamefont {Y.}~\bibnamefont {Tanaka}}, \
  and\ \bibinfo {author} {\bibfnamefont {N.}~\bibnamefont {Nagaosa}},\ }\href
  {\doibase 10.1103/PhysRevB.93.064416} {\bibfield  {journal} {\bibinfo
  {journal} {Phys. Rev. B}\ }\textbf {\bibinfo {volume} {93}},\ \bibinfo
  {pages} {064416} (\bibinfo {year} {2016})}\BibitemShut {NoStop}%
\bibitem [{\citenamefont {Laiho}\ \emph
  {et~al.}(2003{\natexlab{b}})\citenamefont {Laiho}, \citenamefont
  {L\"ahderanta}, \citenamefont {Sonin},\ and\ \citenamefont
  {Traito}}]{laiho_penetration_2003}%
  \BibitemOpen
  \bibfield  {author} {\bibinfo {author} {\bibfnamefont {R.}~\bibnamefont
  {Laiho}}, \bibinfo {author} {\bibfnamefont {E.}~\bibnamefont {L\"ahderanta}},
  \bibinfo {author} {\bibfnamefont {E.~B.}\ \bibnamefont {Sonin}}, \ and\
  \bibinfo {author} {\bibfnamefont {K.~B.}\ \bibnamefont {Traito}},\ }\href
  {\doibase 10.1103/PhysRevB.67.144522} {\bibfield  {journal} {\bibinfo
  {journal} {Phys. Rev. B}\ }\textbf {\bibinfo {volume} {67}},\ \bibinfo
  {pages} {144522} (\bibinfo {year} {2003}{\natexlab{b}})}\BibitemShut
  {NoStop}%
\end{thebibliography}%

\end{document}